\documentclass[%
superscriptaddress,notitlepage,
aps,
jmp,%
amsmath,amssymb,
preprint,%
pra,letter,%
]{revtex4-1}
\usepackage{graphicx}% Include figure files
\usepackage{subfigure}
\usepackage{dcolumn}% Align table columns on decimal point
\usepackage{bm}% bold math
\usepackage[mathlines]{lineno}% Enable numbering of text and display math
\usepackage{array}
\usepackage{multirow}
\usepackage{diagbox}
\usepackage{appendix}
\usepackage{bm,color,bbm}

\newcommand{\blk}{\color{black}}

\usepackage{float}
\newcommand{\ket}[1]{ |{#1} \rangle}
\newcolumntype{.}{D{.}{.}{-1}}
\usepackage{amsopn}
\usepackage[colorlinks,linkcolor=black,anchorcolor=black,  urlcolor=black, citecolor=blue]{hyperref}
\usepackage{epstopdf}

\begin{document}

\title{Experimental Quantum Switching for Exponentially Superior Quantum Communication Complexity}

\author{Kejin Wei}
\thanks{These authors contributed equally to this work.}
\affiliation{Shanghai Branch, Hefei National Laboratory for Physical Sciences at Microscale and Department of Modern Physics, University of Science and Technology of China, Shanghai, 201315, China}
\affiliation{CAS Center for Excellence and Synergetic Innovation Center in Quantum Information and Quantum Physics, University of Science and Technology of China, Shanghai 201315, P. R. China}
%\affiliation{Guangxi Key Laboratory for Relativistic Astrophysics, School of Physics Science and Technology, Guangxi University, Nanning 530004, China}

\author{Nora Tischler}
\thanks{These authors contributed equally to this work.}
\affiliation{Centre for Quantum Dynamics, Griffith University, Brisbane, QLD 4111, Australia}

\author{Si-Ran Zhao}
\affiliation{Shanghai Branch, Hefei National Laboratory for Physical Sciences at Microscale and Department of Modern Physics, University of Science and Technology of China, Shanghai, 201315, China}
\affiliation{CAS Center for Excellence and Synergetic Innovation Center in Quantum Information and Quantum Physics, University of Science and Technology of China, Shanghai 201315, P. R. China}

\author{Yu-Huai Li}
\affiliation{Shanghai Branch, Hefei National Laboratory for Physical Sciences at Microscale and Department of Modern Physics, University of Science and Technology of China, Shanghai, 201315, China}
\affiliation{CAS Center for Excellence and Synergetic Innovation Center in Quantum Information and Quantum Physics, University of Science and Technology of China, Shanghai 201315, P. R. China}

\author{Juan Miguel Arrazola}
\affiliation{Xanadu, 372 Richmond Street W, Toronto, Ontario M5V 1X6, Canada}
\affiliation{Centre for Quantum Technologies, National University of Singapore, 3 Science Drive 2, Singapore 117543}

\author{Yang Liu}
\affiliation{Shanghai Branch, Hefei National Laboratory for Physical Sciences at Microscale and Department of Modern Physics, University of Science and Technology of China, Shanghai, 201315, China}
\affiliation{CAS Center for Excellence and Synergetic Innovation Center in Quantum Information and Quantum Physics, University of Science and Technology of China, Shanghai 201315, China}

\author{Weijun Zhang}
\affiliation{State Key Laboratory of Functional Materials for Informatics, Shanghai Institute of Microsystem and Information Technology, Chinese Academy of Sciences, Shanghai 200050, China}

\author{Hao Li}
\affiliation{State Key Laboratory of Functional Materials for Informatics, Shanghai Institute of Microsystem and Information Technology, Chinese Academy of Sciences, Shanghai 200050, China}

\author{Lixing You}
\affiliation{State Key Laboratory of Functional Materials for Informatics, Shanghai Institute of Microsystem and Information Technology, Chinese Academy of Sciences, Shanghai 200050, China}

\author{Zhen Wang}
\affiliation{State Key Laboratory of Functional Materials for Informatics, Shanghai Institute of Microsystem and Information Technology, Chinese Academy of Sciences, Shanghai 200050, China}

\author{Yu-Ao Chen}
\affiliation{Shanghai Branch, Hefei National Laboratory for Physical Sciences at Microscale and Department of Modern Physics, University of Science and Technology of China, Shanghai, 201315, China}
\affiliation{CAS Center for Excellence and Synergetic Innovation Center in Quantum Information and Quantum Physics, University of Science and Technology of China, Shanghai 201315, P. R. China}

\author{Barry C. Sanders}
\affiliation{Shanghai Branch, Hefei National Laboratory for Physical Sciences at Microscale and Department of Modern Physics, University of Science and Technology of China, Shanghai, 201315, China}
\affiliation{Institute for Quantum Science and Technology, University of Calgary, Alberta T2N 1N4, Canada}
\affiliation{Program in Quantum Information Science, Canadian Institute for Advanced Research, Toronto, Ontario M5G 1M1, Canada} 	

\author{Qiang Zhang}
\affiliation{Shanghai Branch, Hefei National Laboratory for Physical Sciences at Microscale and Department of Modern Physics, University of Science and Technology of China, Shanghai, 201315, China}
\affiliation{CAS Center for Excellence and Synergetic Innovation Center in Quantum Information and Quantum Physics, University of Science and Technology of China, Shanghai 201315, P. R. China}

\author{Geoff J. Pryde}
% \email{g.pryde@griffith.edu.au}
\affiliation{Centre for Quantum Dynamics, Griffith University, Brisbane, QLD 4111, Australia}

\author{Feihu Xu}		
% \email{feihuxu@ustc.edu.cn}
\affiliation{Shanghai Branch, Hefei National Laboratory for Physical Sciences at Microscale and Department of Modern Physics, University of Science and Technology of China, Shanghai, 201315, China}
\affiliation{CAS Center for Excellence and Synergetic Innovation Center in Quantum Information and Quantum Physics, University of Science and Technology of China, Shanghai 201315, P. R. China}

\author{Jian-Wei Pan}
% \email{pan@ustc.edu.cn}
\affiliation{Shanghai Branch, Hefei National Laboratory for Physical Sciences at Microscale and Department of Modern Physics, University of Science and Technology of China, Shanghai, 201315, China}
\affiliation{CAS Center for Excellence and Synergetic Innovation Center in Quantum Information and Quantum Physics, University of Science and Technology of China, Shanghai 201315, P. R. China}

\begin{abstract}
	Finding exponential separation between quantum and classical information tasks is like striking gold in quantum information research. Such an advantage is believed to hold for quantum computing but is  proven for quantum communication complexity. Recently, a novel quantum resource called the quantum switch---which creates a coherent superposition of the causal order of events, known as quantum causality---has \blk been harnessed theoretically in a new protocol providing provable exponential separation. We  experimentally demonstrate such an advantage by realizing a superposition of communication directions for a two-party distributed computation. Our photonic demonstration employs $d$-dimensional quantum systems, qudits, up to $d=2^{13}$ dimensions and demonstrates a communication complexity advantage, requiring less than $(0.696 \pm 0.006)$ times the communication of any causally ordered protocol. These results elucidate the crucial role of the coherence of communication direction in achieving the  exponential separation for the one-way processing task, and open a new path for experimentally exploring the fundamentals and applications of  advanced features of indefinite causal structures.
\end{abstract}
\maketitle

Computation by separated parties  with minimal communication is the focus of communication complexity, which has applications to distributed computing, very-large-scale integration, streaming algorithms, and more~\cite{Buhrman2010}. For quantum information, communication complexity is especially exciting as exponential quantum-classical gaps can be proven~\cite{Yao1979,Brassard2003,Raz1999,Buhrman2001,Babai2004,Gavinsky2007,Regev2011,Arrazola2014a}. By contrast, exponential quantum-classical gaps for computation tasks such as factorization~\cite{shor1999polynomial} depend on the best-known classical algorithm, and thus are strongly believed but not rigorously proven. Experimentally, quantum communication  complexity has been studied in proof of principle for the quantum fingerprinting protocol  \cite{Barry,Xu2015,Guan2016} and beyond \cite{Buhrman2010,du2006experimental,trojek2005experimental}.

% Exponential separations are important to quantum information science as they demonstrate  its stark advantage over its classical counterpart.

The quantum switch provides a new communication complexity tool that leads to another instance of exponential quantum advantage~\cite{Guerin2016}. The quantum switch is a device where a control qubit determines the order in which two transformations are performed on a target system~\cite{Oreshkov2012,Chiribella2013}. When the control is in a superposition of logical states, the order of the operations is causally indefinite; i.e., there is a superposition of the ordering of target operations. The quantum switch has broad relevance in the context of quantum causality~\cite{Brukner2014} including applications to studies of quantum gravity~\cite{Hardy2007,Zych2017,Brukner2014,Rubino}, communication complexity~\cite{Feix2015,Guerin2016}, witnessing causality~\cite{Oreshkov2012,Araujo2015,Branciard2016a,Rubino2017,Goswami2018,Castro2018} and deciding whether a given indefinite causal order is physical~\cite{Araujo2017,oreshkov2018whereabouts}. In quantum computing, the quantum switch can reduce the query complexity for some tasks compared to causally ordered protocols~\cite{Chiribella2013,Araujo2014} --- this advantage has been demonstrated for single-qubit control and single-qubit target circuits~\cite{Procopio2015}. In quantum communication, the quantum switch enhances the communication rate beyond the limits of conventional quantum Shannon theory~\cite{Ebler2018,goswami2018communicating}.

In this letter, we demonstrate a quantum-switch-based exponentially enhanced quantum-communication protocol, modified from a recent theory proposal~\cite{Guerin2016}. We implement a single-qubit-controlled switch acting on a target Hilbert space of~$d$ dimensions realized as a qudit~\cite{BdGS02,PZ88}, which simplifies the experiment compared to using multiple qubits while maintaining the Hilbert-space dimension. Unlike past experimental qubit implementations~\cite{Procopio2015,Rubino2017,Goswami2018,Rubino,goswami2018communicating}, our experiment provides the first demonstration of the quantum switch involving a qudit target system.

%Our experiment thus provides a proof-of-principle demonstration of quantum switching for achieving exponentially superior performance in a quantum-communication-based task.
% Our experiment demonstrates an exponential communication complexity advantage over any causally ordered protocol.

\begin{figure}[!hbt]
	\centering
	\includegraphics[width=0.5\linewidth]{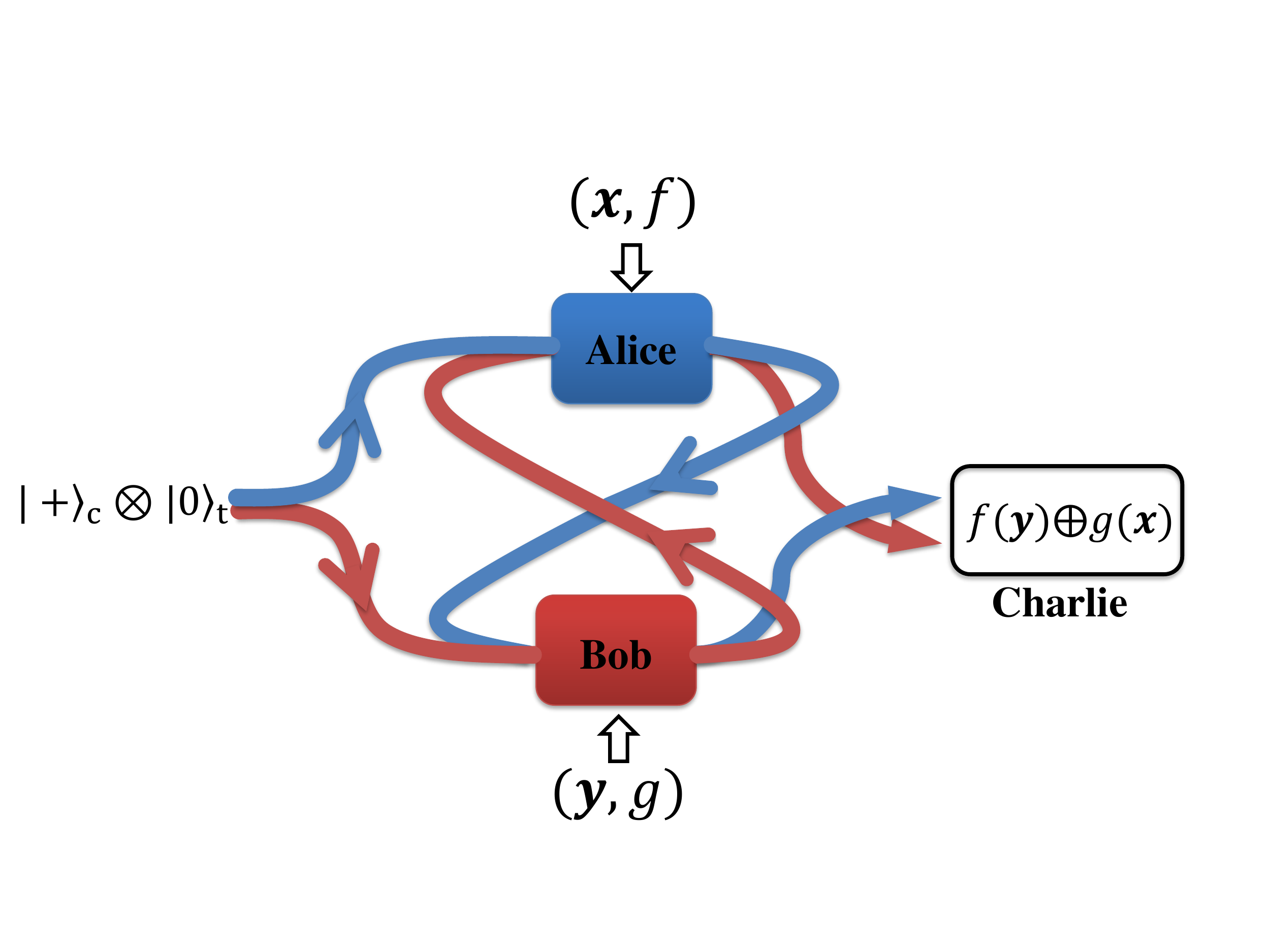}
	\caption{ Schematic illustration of the exchange evaluation game. Alice and Bob receive inputs $(\bm{x},f)$ and $(\bm{y},g)$, respectively. There is another party, Charlie, who needs to compute the exchange evaluation function $\operatorname{EE}(\bm{x},f,\bm{y},g)=f(\bm{y})\oplus g(\bm{x})$. Under the condition of one-way communication, for a causally ordered protocol, Alice and Bob can transmit information either from Alice to Bob, then Charlie (blue arrow) or from Bob to Alice, then Charlie (red arrow). Using a quantum switch,  a control qubit in the state $\ket{+}_{\rm c}=1/\sqrt{2}\left(|0\rangle_{\rm c}+|1\rangle_{\rm c}\right)$ coherently controls the communication direction of the target system  $\ket0_{\rm t}$. Charlie computes the exchange evaluation function by measuring the control qubit. As a result, \blk they solve the game with exponential advantage in the scaling of the required communication compared to any causally ordered protocol.
		\label{exchange}}
\end{figure}

The communication task that we consider is known as the exchange evaluation (EE) game, illustrated in Fig.~\ref{exchange}.
For bit strings $\bm{x},\bm{y}\in\{0,1\}^n$
and Boolean functions $f,g:\{0,1\}^n\to\{0,1\}$,
Alice and Bob receive inputs $(\bm{x},f)$ and $(\bm{y},g)$, respectively,
such that $f(\bm{0})=g(\bm{0})=0$,  with~$\mathbf{0}$  being  the all-zero string.
Another party, Charlie,
must compute the exchange evaluation function
$\operatorname{EE}(\bm{x},f,\bm{y},g)=f(\bm{y})\oplus g(\bm{x})$
subject to the one-way communication condition
that Alice and Bob can each send a system to the outside environment only once. They
are thus forbidden to communicate back and forth by sending quantum or classical systems to each other at different time steps.

Gu\'{e}rin $et~al.$\ proposed a protocol that uses the superposition of the communication direction between Alice and Bob to solve the EE game with unit success probability, using an amount of communication that scales exponentially better with the input bit string size $n$ compared to any causally ordered protocol~\cite{Guerin2016}. They cast the task in terms of the problem of deciding whether two unitary operations, which depend on the inputs of Alice and Bob, commute or anticommute. This constitutes precisely the question that can be solved efficiently with the quantum switch. Importantly, the communication complexity advantage is not limited to the ideal case of deterministic answers, which are unachievable in any real experiment; it also survives for the more practical bounded-error case. Despite the elegance of the protocol in theory, it is challenging in experiment because the protocol typically requires the manipulation of more than 10 entangled qubits to enter the regime where the causally indefinite advantage starts.

We implement a variant of the protocol that renders it experimentally feasible with current technology (see Appendix~\ref{Compari} and~\ref{Practical}). In our protocol,  we use a quantum switch where one qubit acts as the control, and the target system is a $d$-dimensional qudit~\cite{BdGS02} with $d=2^{n+1}$, thus making it equivalent to $(n+1)$ entangled qubits. We implement both systems with a single photon by encoding the control system in the path degree of freedom and the target system in the arrival time of the photon, which is divided into a set of $2^{n+1}$ time bins forming the basis $\{\ket{z};z\in\{0,1,\ldots,2^{n+1}-1\}\}$. Consistent with the notation for~$z$, we write~$x$ and~$y$ as the unary representations of the bit strings~$\bm x$ and~$\bm y$, respectively, noting that the maximum value they can take is $2^n-1$. To account for the different dimensionality of the target system and these bit strings (See Appendix~\ref{Compari}), we trivially augment the functions $f$ and $g$ as $f(z)=g(z)=0$ for $z\in \{ 2^n, 2^n+1,... ,2^{n+1}-1 \}$.

Our protocol tests whether two unitary operators~$U_A$ and~$U_B$ commute or anticommute when applied to a particular target state. The relationship between this task and the communication complexity one (evaluating $f({y})\oplus g({x})$) can be seen by representing $U_A$ and~$U_B$ in the
unitary-operator basis given by the generalized Pauli operators~\cite{BdGS02,PZ88} $X_d\ket{z}=\ket{z+1\mod d}$
and
$Z_d\ket{z}=\exp(2\pi{\rm i}z/d)\ket{z}$
satisfying
$Z_dX_d=\exp(2\pi{\rm i}/d)X_dZ_d$.
In this basis,
we write $D_d(f)\ket{z}=Z^{f(z)d/2z}_d\ket{z}$
and define $U_A\equiv X_d^xD_d(f)$ and $U_B \equiv X_d^yD_d(g)$. Thus,
\begin{equation}
\left[U_A,U_B\right]\ket0_{\rm t}	=0\textrm{ if }f({y})\oplus g({x})=0,\;
\left\{U_A,U_B\right\}\ket0_{\rm t}=0\textrm{ if }f({y})\oplus g({x})=1;
\end{equation}
where t denotes the target system.

 The quantum switch can be used to solve this problem as follows (see Fig.~\ref*{exchange}). \blk
The control qubit is prepared in the state
$\ket{+}_{\rm c}=1/\sqrt{2}\left(|0\rangle_{\rm c}+|1\rangle_{\rm c}\right)$, and the target system is initialized in $\ket0_{\rm t}$. After applying the unitary operators $U_A$ and $U_B$ to the target system in the order determined by the control qubit, a Hadamard gate is applied to the control qubit, with the resulting state of the overall system being $\frac{1}{2}(|0\rangle_{\rm c}\otimes\{U_A,U_B\}\ket0_{\rm t} -|1\rangle_{\rm c}\otimes[U_A,U_B]| {0}\rangle_{\rm t})$.
Then, a measurement of the control qubit by Charlie in the computational basis reveals whether $f({y})\oplus g({x})=0$ or $f(y)\oplus g(x)=1$, which completes the communication-complexity task.

To implement the protocol, the state preparation and the Hadamard gate for the control qubit encoded in the path degree of freedom are realized using a beam splitter (BS). The target qubit initialization in $|0\rangle_{\rm t}$ is equivalent to the photon starting in the first time bin. The transformations $X({x}) =X_d^x $ and $X({y})=X_d^y$ are implemented as delays by~$x$ time bins in Alice's laboratory and~$y$ time bins in Bob's laboratory, respectively. Furthermore, $D_d(f)$ is implemented in Alice's laboratory using a phase modulator that applies the appropriate phase (0 if $f(z)=0$ or $\pi$ if $f(z)=1$) to each time bin, with a corresponding implementation for $D_d(g)$ in Bob's laboratory.

\begin{figure*}[!hbt]
	\centering
	\includegraphics[width=0.9\linewidth]{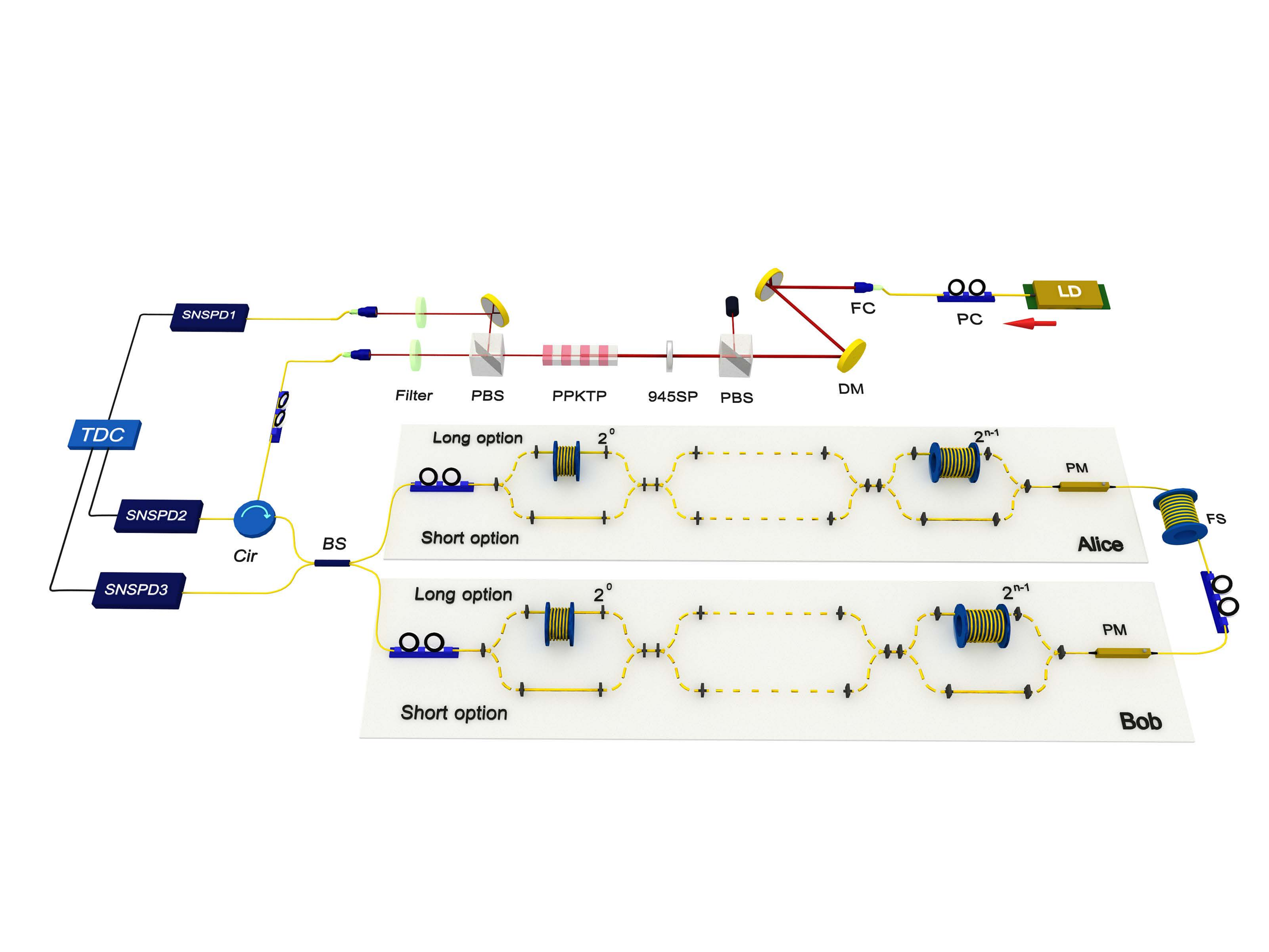}
	\caption{ Experimental setup of the causally indefinite protocol. The pump light is emitted by a 780-nm continuous-wave laser diode (LD). After passing through a 945-nm short-pass filter (945SP) used to remove the residual long wavelengths from outside fluorescence, the pump light is focused into a 10 mm, periodically poled potassium titanyl phosphate (PPKTP) crystal to convert pump photons into pairs of orthogonally polarized photons at a wavelength of 1560 nm.  The down-converted photons are separated by a polarizing beam splitter (PBS), and the residual pump photons are removed by two filters. One of the down-converted photons is sent to a circulator (Cir) and then enters the Sagnac loop  via a beam splitter (BS). To herald the presence of this photon, the other one is directly detected by a high-quality superconducting nanowire single-photon detector (SNSPD1). In the Sagnac loop, Alice and Bob each possess a variable delay and a phase modulator (PM) to implement their unitary transformations. For each of the $n$ segments, either the long option or the short option is chosen, depending on whether the corresponding digit of~$x$ in binary represention is 0 or 1. A fiber spool (FS) with a length of 7 km is used to store the pulse train. The interference results are monitored by SNSPD2 and SNSPD3. Abbreviations of other components: PC, polarization controller; FC, fiber coupler; DM, dichroic mirror; TDC, time-to-digital converter.
		\label{fig1}}
\end{figure*}

The experiment utilities a fiber-based Sagnac interferometer, shown in Fig.\ \ref{fig1}. A heralded single photon, generated by type-\uppercase\expandafter{\romannumeral2} spontaneous parametric down-conversion, passes a circulator and then enters the Sagnac loop, via a BS, in a superposition of two directions. Alice and Bob each possess a variable delay line and a time-dependent phase modulator  to realize the unitary transformations $U_A$ and $U_B$, respectively. They each implement the transformation once on the component of the photon that passes through them after entering the loop, and once more on the component that passes through them before exiting the loop. Depending on the binary answer to the EE game, the photon interferes such that it exits at one or the other of the two interferometer outputs. The outputs are monitored by two high-quality superconducting nanowire single-photon detectors (SNSPDs). The detection events are registered using a time-to-digital converter.

In the experimental implementation of the communication complexity protocol, the amount of communication and  the \blk error probability are to be minimized. These goals lead to three main technological challenges (see Appendix~\ref{C}). (i) Precise timing is required to ensure that the correct phase is imparted at the second station. We use a sequence of home-made fiber segments to create the $2^n$ choices of delays for each of Alice and Bob with a high accuracy, and a 25-GHz arbitrary waveform generator, triggered by the heralding signal, in such a way that the modulation timing accurately coincides with the time bins   (of length 2~ns) of the heralded photons.   (ii)   To minimize noise, we use two SNSPDs with ultralow dark-count rates below $10$ Hz, and perform careful temperature control and polarization control to maximize the interference visibility. (iii) It requires the optimization of the efficiency throughout the setup, to minimize the amount of required communication. For example, we need a loss $<15.94$ dB to beat the classical definite protocol with $n=10$. To this end, we choose suitable pump and detection beam waists for the down-conversion process, use custom-made low-loss components in the Sagnac loop, and detect the photons with high-efficiency SNSPDs.

\begin{figure}[!hbt]
	\centering
	\includegraphics[width=0.5\linewidth]{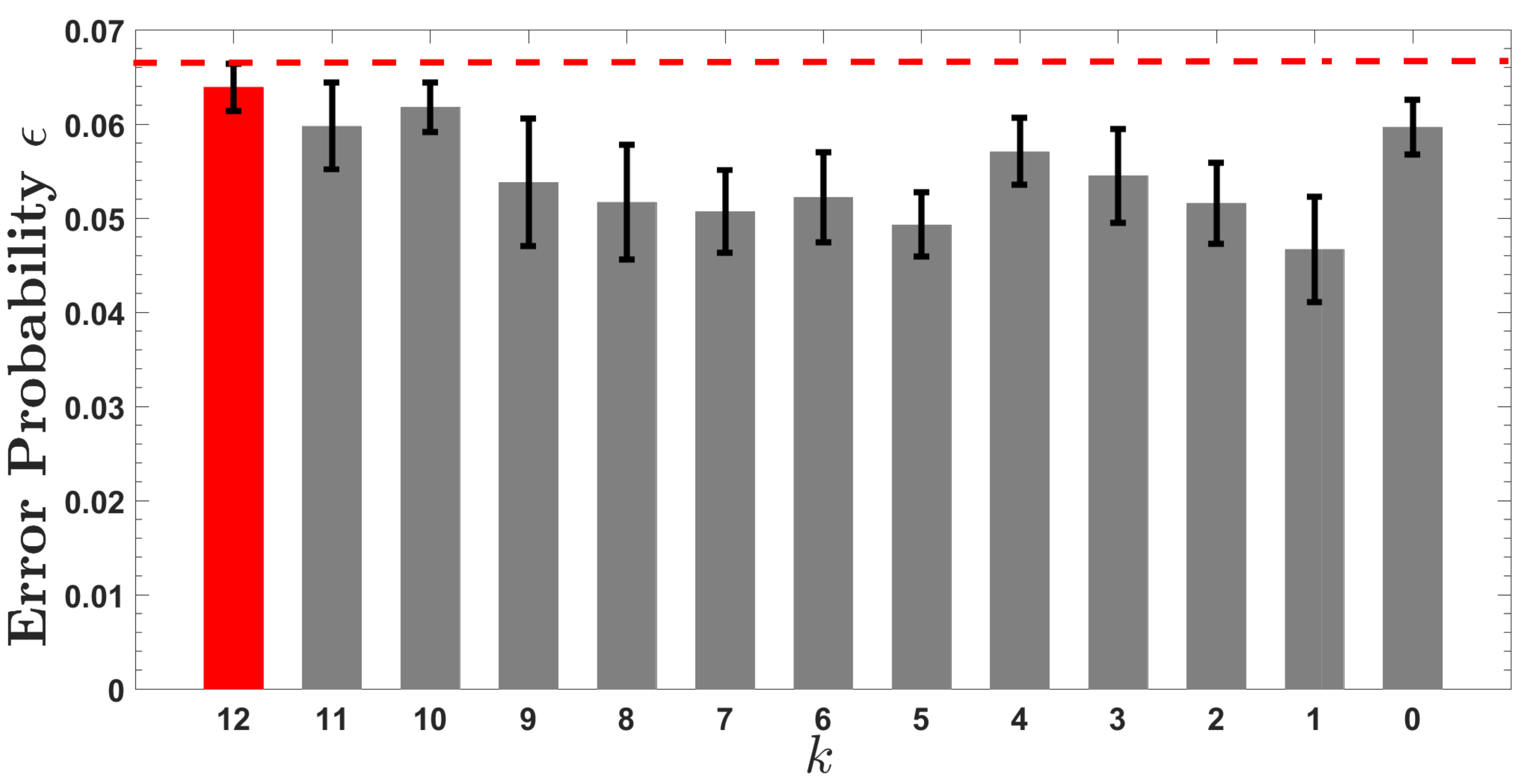}
	\caption{(Color online) Error probability for the system size $n=12$. The value $k=12$ denotes the worst case of inputs $x=y=2^n-1$. The other values of $k\in{\{0,1,\ldots,11\}}$ correspond to Alice's $(k+1)$-th bit being flipped from 1 to 0. The red dashed line shows the worst error probability including the uncertainty. The error bars indicate 1 standard deviation.
		\label{error12bits}}
\end{figure}

In our experiment, we determine the error probability and the transmitted information for each $n\in{\{9,10,11,12\}}$. We record two-photon  coincidence counts (between SNSPD1 and either SNSPD2 or SNSPD3) ( Fig.~{\ref{fig1}}). We estimate the error probability $\epsilon$ (i.e. the probability of an incorrect result) focusing on  two important cases: the worst case, where both Alice and Bob have the largest input $x=y=2^n-1$ (corresponding to the largest delays and system loss); and  the one-bit-different case where Alice changes the choice of the $(k+1)$-th segment, $k\in{\{0,1,2...,n-1\}}$, while Bob holds the largest input. The goal of testing the second case is to verify that the delay of each segment is well made, so that Alice has the ability to implement any of the $2^n$ possible delays. Figure\ \ref{error12bits} shows the detailed experimental error probability for $n=12$, where we obtained a value of $\epsilon= 0.0638\pm0.0025$ for the worst case (red bar). As expected, for other values of $k$, the error probability is below that of the worst case. For other values of $n$, see the Appendix~\ref{detailed experimental result}.

\begin{figure}[!hbt]
	\centering
	\subfigure[]{\label{results1}
		\begin{minipage}[t]{0.5\linewidth}
			\centering
			\includegraphics[width=1\linewidth]{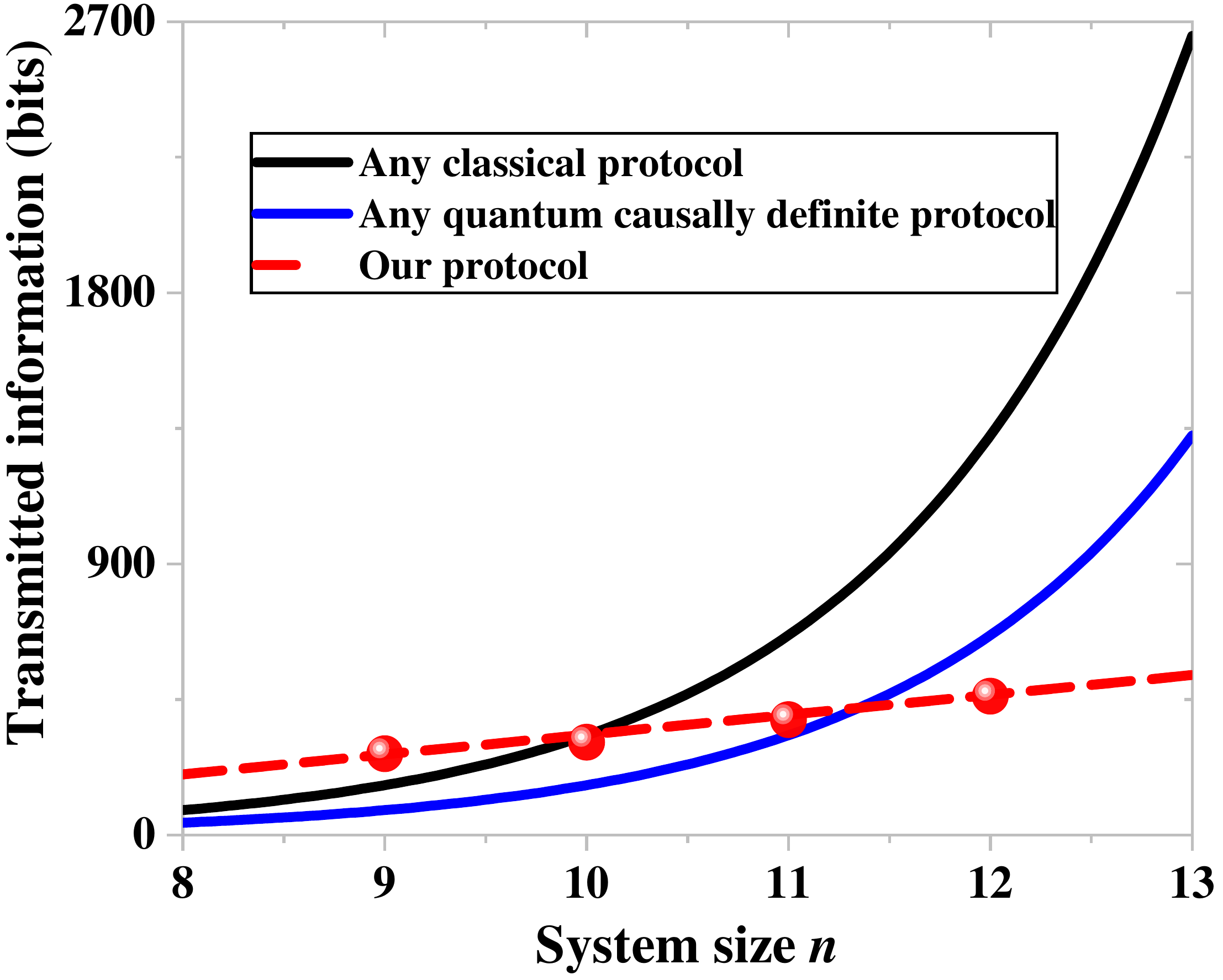}
	\end{minipage}}
	\subfigure[]{\label{results2}
		\begin{minipage}[t]{0.5\linewidth}
			\centering
			\includegraphics[width=1\linewidth]{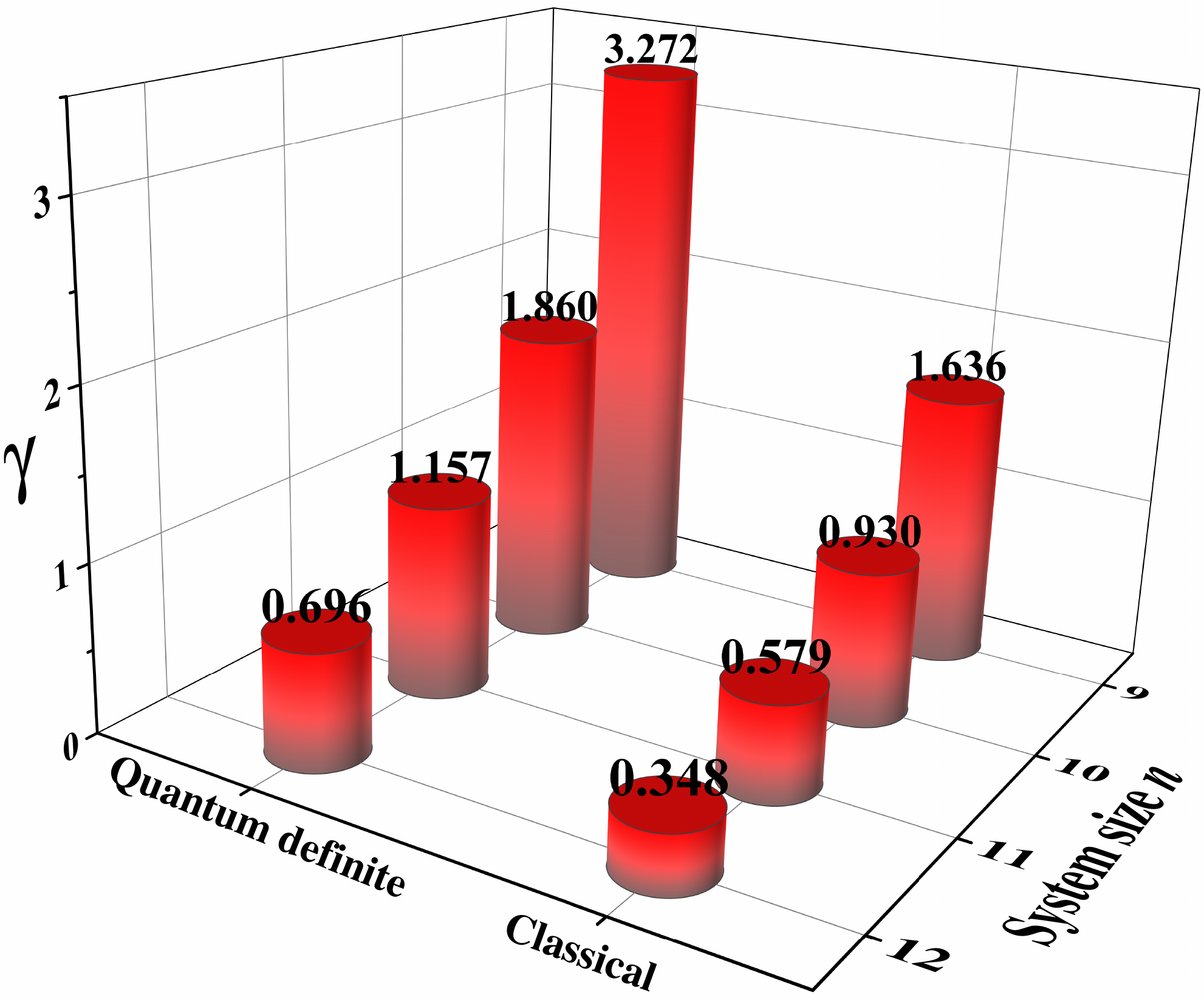}
	\end{minipage}}
	\caption{ (a) Comparison of the transmitted information. The red points show our experimental results for different system sizes. The red dashed line is the best fit straight line for the four data points.  The error bars are too small to be seen.
		For $n=12$ our experimental result clearly beats both classical protocols and quantum causally definite protocols. (b) The ratio $\gamma$ of the transmitted information $Q$ of our causally indefinite protocol, and the lower bound on the transmitted information $C$ for a given causally ordered protocol.}\label{information}
	
\end{figure}

Figure\ \ref{results1} shows the experimental transmitted information required to complete the task \blk for different system sizes. We compare our protocol with a bound for classical protocols (black solid curve) and a bound for quantum causally definite protocols (blue solid curve). A lower bound on the transmitted information in any classical protocol, allowing for a bounded error probability of $\epsilon$, is $[1 - H(\epsilon )]{2^{n - 1}}$, where $H(\epsilon)$ is the binary entropy. Any quantum causally definite protocol needs to transmit at least $[1 - H(\epsilon )]{2^{n - 2}}$, half as much as classical protocols~\cite{Guerin2016}. We use the worst-case $\epsilon$ ($= 0.0663$) to calculate the minimum transmitted information of classical and quantum causally definite protocols, because it gives the most stringent bound to beat. For each $n$, the transmitted information of our protocol is estimated to be $(n+2)/\eta_n$, where  $\eta_n$ ($n\in{\{9,10,11,12\}}$)  denotes the worst-case overall system efficiency for the given system size. Figure\ \ref{results1} indicates that, with increasing system size $n$, the causally indefinite protocol provides an exponential advantage in the scaling of transmitted information. In particular, for $n=12$, the experimental results clearly beat all classical protocols as well as all quantum causally definite protocols.

To further quantify our results, we define $\gamma$ as the ratio between  the transmitted information $Q$ of our causally indefinite protocol, and the minimum transmitted information $C$ of a given type of causally ordered protocol (either classical or quantum causally definite): $\gamma:=Q/C$. In Fig.\ \ref{results2}, we plot the experimental results for $\gamma$ as a function of different protocols and different input sizes. Figure \ \ref{results2} shows that our protocol requires the transmission of only $(34.8 \pm 0.3)\%$ of the information of any classical protocol and $(69.6 \pm 0.6)\%$ of the information of any causally definite quantum protocol.

 Our experimental results show that we have demonstrated the EE
game, even taking into account small experimental imperfections.  First, we measured the second-order correlation function at zero delay of our heralded single-photon source, and obtained  $g^{(2)}(0)=(3.27\pm23.5)\times 10^{-5}$, where the error bars represent 1 standard deviation. The value is close to 0, which ensures the condition of one-way communication in EE game, as discussed in \cite{Guerin2016,Procopio2015}(see Appendix~\ref{Measurement second}).  Second, the random choice of the delay and the phase modulation were carefully characterized  (see Tab.~S1 in~\ref{C}), which guaranteed the implementation of arbitrary bit inputs ($\bm{x}$,$\bm{y}$) and functions ($f$, $g$), as required by the EE game.
Third, although the error probability is rising with increasing system size, by careful optimizing the setup, the results of Fig.~\ref{error12bits} show that the error probability is below a tolerable range. Such a small error is sufficient to realize the EE game with causally indefinite quantum protocol. Fourth, though the optical loss is exponential growth with the increasing system size, the results of Fig.~\ref{information} prove that with system sizes of 13 bits, we can access the regime where our causally indefinite quantum protocol outperforms any causally definite protocol.

\blk Overall, we have experimentally demonstrated that the superposition of communication direction can provide an exponential reduction of communication complexity. We achieve this through the use of a high-dimensional degree of freedom, the photon arrival time. Although our temporal qudit encoding is not efficiently scalable to very large $n$, our experiment elucidates the crucial role of coherence in achieving the exponential separation between quantum and classical communication in a distributed processing task. Decoherence would destroy the superposition of communication directions---reverting to a situation where it is necessary to transmit the function itself through the one-way channel to achieve deterministic two-party function evaluation. Our results establish that the quantum switch can serve as a powerful resource in quantum communication, which may provide a new paradigm of Shannon theory, where the order of the communication channels can be in a quantum superposition, and therefore strongly motivates further experimental development with qudit or other encodings.

\begin{acknowledgments}
	This work was supported by the National Key R\&D Program of China (Grant No. SQ2018YFB0504303, SQ2018YFB050101), the National Natural Science Foundation of China (Grants No. 61771443 and No. 61705048), the Chinese Academy of Sciences, the Thousand Young Talent Program of China, Shanghai Science and Technology Development Funds (Grant No. 18JC1414700), and the Australian Research Council (Grant No. DP160101911). The authors would like to thank Norbert L\"{u}tkenhaus, Farzad Ghafari, Jian-Yu Guan, and Cheng Wu for helpful discussions.
\end{acknowledgments}
\newpage
\appendix
\section{Comparison of our protocol with that of Ref.\ \cite{Guerin2016}}\label{Compari}

\begin{figure}[!h]
	\centering
	\includegraphics[width=0.6\linewidth]{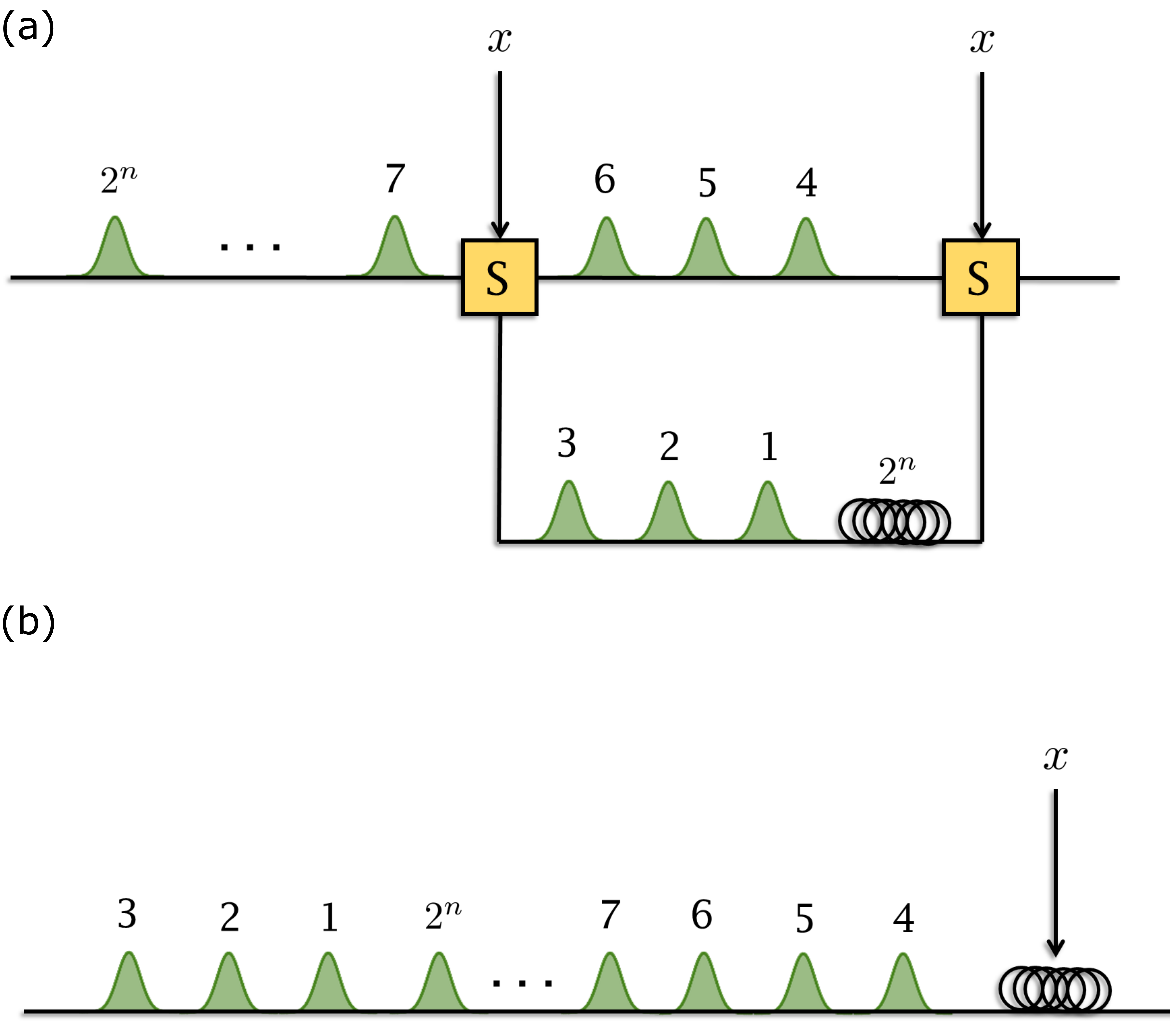}
	\caption{Linear optic circuit for performing an alternative version of the transformation $X(x)$, such that the number of qubits in the target system is reduced to $n$. (a) The $2^{n}$ incoming modes pass through an active switch S which sends each mode either straight through or sends it to a delay line. The delay is equivalent to $2^{n}$ time bins. The operation of the switch depends on $x$ and it is such that only the first $\left(2^{n}-x\right)$ modes are sent through the delay. Here we have used $x=2^{n}-3$ as an example. A second switch then ensures that all modes exit in the same line. (b) After the previous step, the modes have been permuted as desired and an further delay equal to $x$ time bins must be added to ensure that the leading mode reaches the other party at the same time, regardless of the value of $x$.}
	
	\label{fig:permutation}
\end{figure}

Our protocol differs slightly from the original scheme proposed in Ref.\ \cite{Guerin2016}. The unitary transformation $X(\bm{x})$ was originally proposed in terms of modulo-two addition, as $X(\bm{x})=\sum_{\bm{z}\in\{0,1\}^{n}}|\bm{z}\oplus\bm{x}\rangle\langle\bm{\bm{z}}|$. Our definition of  $X(x)=\sum_{z}|(z+x)~ \mathrm{mod}~ 2^{n+1} \rangle\langle z|$ retains the key property of $[X(x),X(y)]=0 ~\forall x~and~y $, while being experimentally much more readily realizable. Note that the fact that our target system is equivalent to $n+1$ qubits as opposed to the $n$ qubits of Ref.\ \cite{Guerin2016} is not a fundamental consequence of the time-bin implementation. One could also have the target system corresponding to $n$ qubits, and use the transformation $X(x)=\sum_{z=0}^{2^{n}-1}|(z+x) ~\mathrm{mod}~ 2^n \rangle\langle z|$. Such a permutation would require using fast switches in addition to the delays, as shown in Fig.\ \ref{fig:permutation}.
\section{Practical considerations}\label{Practical}
Our encoding of the whole multi-qubit system in a single photon entails both benefits and drawbacks. It is beneficial given the presence of probabilistic photon loss, an unavoidable experimental imperfection. Efficiency is a crucial factor in demonstrating the advantage provided by the quantum switch, and presents one of the main technical challenges of the experiment. Loss is detrimental because it effectively requires the repetition of the protocol, thereby increasing the communication, while the performance is benchmarked against optimal lossless causally ordered protocols. As a result of our implementation, our quantum system is only subjected to the probabilistic loss associated with one photon. However, the encoding of the target qubits in the photon arrival time also leads to an exponential growth of the number of required time bins as a function of the number of target qubits. This practically limits the ability to scale the protocol to an arbitrarily high value of $n$. Nevertheless, our experiment proves that it is possible to access the regime where our causally indefinite quantum protocol outperforms any causally definite protocol.

In contrast to previous communication complexity experiments on quantum fingerprinting that were based on coherent states \cite{Xu2015,Guan2016}, we use single photons in this experiment. The use of coherent states would arguably violate a rule of the exchange evaluation game, namely that Alice and Bob are each only allowed to receive a system from the outside environment once. As a test of one-way communication, Ref.\cite{Guerin2016} proposed having a (hypothetical) counter in each laboratory that is incremented by one whenever a system enters the laboratory. Our single photon constitutes an indivisible quantum system, so the meaning of ``a system entering" is unambiguous, and the counters would read one at the end of the protocol, in agreement with the rules of the game. However, for the case of a coherent state divided on a beam splitter, each component could be considered a quantum system on its own. In that case, the counters would read two, in violation of the rules of the game. It has recently also been experimentally shown that the use of single photons can enable secure communication with a hidden communication direction \cite{2018Del,2018massa}.

The coherence length of our source is approximately several centimeters, which does not cover the whole switch. For our setup, having a coherence length that covers the whole interferometer is not an option. First, we have an extremely long path in the interferometer and it is extremely challenging to have a narrow-bandwidth single photon source. Second, we use time-bin encoding so our photon must be contained in a time bin for modulation – the coherence length must be shorter than the separation between bins.

It is also worth noting that an indefinite causal order has been formally witnessed elsewhere without the use of an especially narrowband single photon~\cite{Rubino2017}.
\blk

\section{Experimental details}\label{C}
\subsection{Heralded single-photon source}
The photon pairs are produced via type-\uppercase\expandafter{\romannumeral2} spontaneous parametric down-conversion. The pump light is emitted by a 780-nm  continuous-wave laser diode (LD). After passing through a 945-nm short-pass filter (945SP) used to remove the residual long wavelengths from outside fluorescence, the pump light is focused into a 10 mm, periodically poled potassium titanyl phosphate (PPKTP) crystal to convert pump photons into pairs of orthogonally polarized photons at a wavelength of 1560 nm.  The down-converted photons are separated by a polarizing beam splitter (PBS), and the residual pump light is removed by two filters. The down-converted photons are then coupled into single-mode fibers with a 76.9\% coupling efficiency. One of them passes a circulator before entering the Sagnac loop. To herald the presence of the correlated photon, the other one is directly detected by a high-quality superconducting nanowire single-photon detector (SNSPD1) with 80.6\% detection efficiency. Overall, the heralding efficiency is approximately $(62.0\pm 1.0)\%$, including all losses in the photon-pair source setup (but excluding the elements used to implement the communication complexity protocol).

\subsection{Sagnac loop}
The Sagnac loop is depicted in Fig.\ 2 of the main text. A 50:50 beam splitter (BS) is used to create the superposition of communication direction. Alice and Bob each possess a variable delay and a phase modulator to implement their unitary operators based on  their inputs when the photon arrives at their stations. The unitaries consist of phase modulations followed by delays. However, because $f(0) = g(0) = 0$ and the photon is initialized in the first time bin, no phase modulation needs to be implemented at the first station, only the delays. The photon then continues along the Sagnac loop, through a 7 km fiber spool in which the pulse trains are stored, and to the second station. There, Alice and Bob each implement their phase modulation, followed by their delay. Finally, the two paths interfere at the BS and the photon is detected by two high-quality SNSPDs (SNSPD2 with a detection efficiency of 76.5\% and dark count rate of 9.3 Hz, SNSPD3 with a detection efficiency of 72.0\% and a dark count rate of 8.3 Hz). The two-photon coincidences are registered by a time-to-digital converter (TDC). All components in the loop have ultra-low insertion loss to minimize the communication complexity.

\subsection{Dark counts}
The coincidence time window in the causally indefinite protocol is very large. In contrast to typical two-photon coincidence experiments where the coincidence window tends to be a few ns, in the causally indefinite protocol, the heralding photon arrives in a given time bin while the heralded photon can be in one of $2^{n+1}$ time bins. This means that the coincidence time window is $2^{n+2}$ ns for the given 2 ns time bin, $\sim (n+1)/3$ orders of magnitude larger than that in normal two-photon coincidence detection. If there is a dark count during this time (or a subset of this interval when applying gating), it will be erroneously registered as an event. To suppress dark counts due to thermal background radiation, we use two high-quality SNSPDs that use a low-loss bandpass filter.
%\red \textbf{[[Sorry to be a pedant, but what is being suppressed is counts from background light or fluorescence, not (thermal) dark counts. Would it be worth saying this a bit more explicitly?]]} \blk
The SNSPDs can detect photons with an ultra-low dark-count rate of $< 10$ Hz and a high detection efficiency of $>72.5\%$.

\subsection{Losses}
To minimize the experimental communication in the causally indefinite protocol, the experimental setup should have as small a loss as possible. Here, three strategies are applied to minimize the total loss of the setup. First, for the heralded single-photon source, by setting the pump beam waist to $340~\mu\mathrm{m}$, the detection beam waist at the center of the PPKTP crystal to $170~\mu\mathrm{m}$, and by heralding the daughter photons using a high-quality SNSPD (detection efficiency of 80.6\%, dark count rate of 11.0 Hz), we achieve a high heralding efficiency of $(62.0\pm 1.0)\%$, corresponding to 2.07 dB loss in the source. Second, in the Sagnac loop, all components are custom-made to achieve ultra-low loss. As discussed in Section E of this Supplemental Material, instead of using highly lossy optical switches, we implement the delay by joining the different fiber segments via mating sleeves. There, we use different connector types, SC/PC to FC/PC mating sleeves (Thorlabs), which provide a lower loss ($ <0.11$ dB), compared with a typical loss of FC/PC to FC/PC mating sleeves ($\sim0.3$ dB). As a result, for each system size $n$, we implement the delays with extremely low loss. Lastly, for the detection we use two high-quality SNSPDs that have a high detection efficiency of 72.0\% and 76.5\% together with an ultra-low dark count rate of 8.3 Hz and 9.3 Hz, respectively. The overall system loss, excluding the loss of the variable delays in Alice and Bob's stations, is 11.62 dB.

We list all losses in the two sections of different lengths (Tab. \ref{tabledelay}), as well as losses in the two sections of each system size (Tab. \ref{tableresults}). The balancing loss in the beamsplitter is 0.27~dB.
\subsection{Measurement of the second-order correlation function}\label{Measurement second}

We need to demonstrate that a pure single photon are used, as this is  underlying assumption\cite{Guerin2016}. This can be shown by measuring the second-order correlation function at zero delay of our heralded single-photon source, $g^{(2)}(0)$. For an ideal heralded single-photon source, the value of $g^{(2)}(0)$ is 0, which implies that only a single photon is sent to the Sagnac loop each time. To measure $g^{(2)}(0)$, in the same way as for the scheme in Ref.~\cite{2005Ren,2015JIN201547}, we simply steer  one of the outputs of the beam splitter (in Fig. 2 of the main text) to SNSPD2, and the other one to SNSPD3. Since the pump light is emitted by a continuous-wave LD, we register all detection events by the TDC and post-select the coincidence events. $g^{(2)}(0)$ is evaluated with the following equation:

\begin{eqnarray}
{g^2}(0) = \frac{{2\times{C_{123}} \times {C_1}}}{{{{({C_{12}} + {C_{13}})}^2}}}
\end{eqnarray}
where $C_1$ is the single count of SNSPD$_1$; $C_{12}$ and $C_{13}$ are two-fold concidence rate between SNSPD$_1$ and SNSPD$_2$, and between SNSPD$_1$ and SNSPD$_3$, respectively; $C_{123}$ denotes the three-fold coincidence rate of  SNSPD$_1$, SNSPD$_2$ and SNSPD$_3$.

At our pump power of 3~mW,we average the count rates over 5 minutes,  obtaining $g^{(2)}(0)=(3.27\pm23.5)\times 10^{-5}$, where the error bars represent 1 standard deviation, following Poisson statistics. The value is very close to 0, which means that an almost perfect single-photon source is used in our experiment. Hence, similarly to Ref. \cite{Procopio2015,Guerin2016}, using the hypothetical photon-counter scheme where we place a counters at the output ports of Alice's and Bob's station to read the number of uses of channel, we would see that the counters of Alice and Bob are 1, which means that the channel is used only once.

\blk
\subsection{Implementation of the unitary transformations}

\begin{figure}[!hbt]
	\centering
	\includegraphics[width=0.8\linewidth]{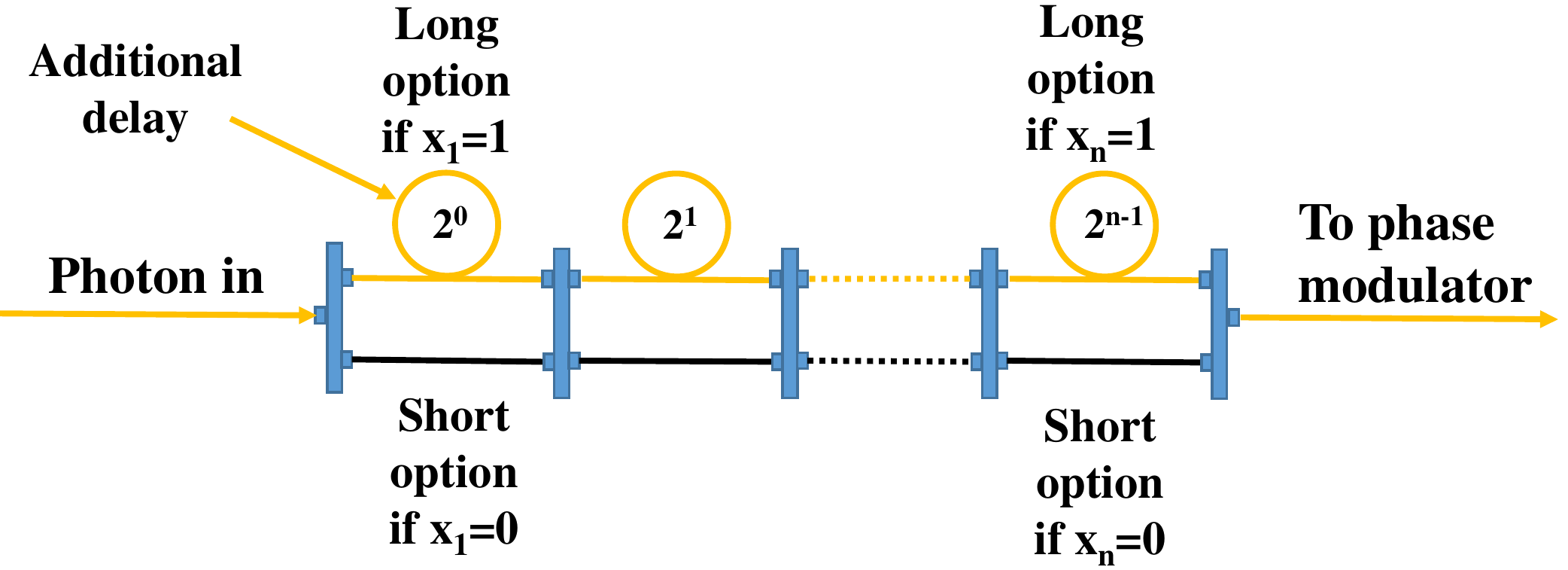}
	\caption{(Color online) Schematic of the implementation of $X(x)$. For each of the $n$ segments, either the short option (black fiber) or the long option (yellow fiber) is chosen, depending on whether the corresponding digit of $x$ in binary representation is $0$ or $1$. For the $k$-th segment, the length of the long option and that of the short option are $t\times(2^{k-1}+1)$ and $t$, respectively, where $t=2$ ns is the pulse width of phase modulation.
	}\label{delay}
\end{figure}

For our protocol, Alice (Bob) needs to realize one of the $2^n$ possible delays, depending on her (his) input bit string, and guarantee that her (his) PM correctly modulates the photon after the delay from Bob (Alice). Therefore, after the delay, the photon pulse must be fully contained in the appropriate time bin. Instead of using highly lossy switches, we implement the variable delays with $n$ segments of fiber, where for the $k$-th segment, there is a choice of two fibers that have a length difference equivalent to a delay of $2^{k-1}$ time bins, $k\in \{{1,2,\ldots,n}\}$, as illustrated in Fig. \ref{delay}. We use a 25 GHz arbitrary waveform generator (Tektronix70002), which is triggered by the heralding signal, to generate a modulation pulse sequence with a pulse width of 2 ns. The rising edge and the falling edge of the pulse are $<100$ ps. The delay between the heralding signal and the modulation pulse is accurately controlled such that the heralded photon is positioned in the middle of the first time bin.

\begin{figure}[!hbt]
	\centering
	\includegraphics[width=0.6\linewidth]{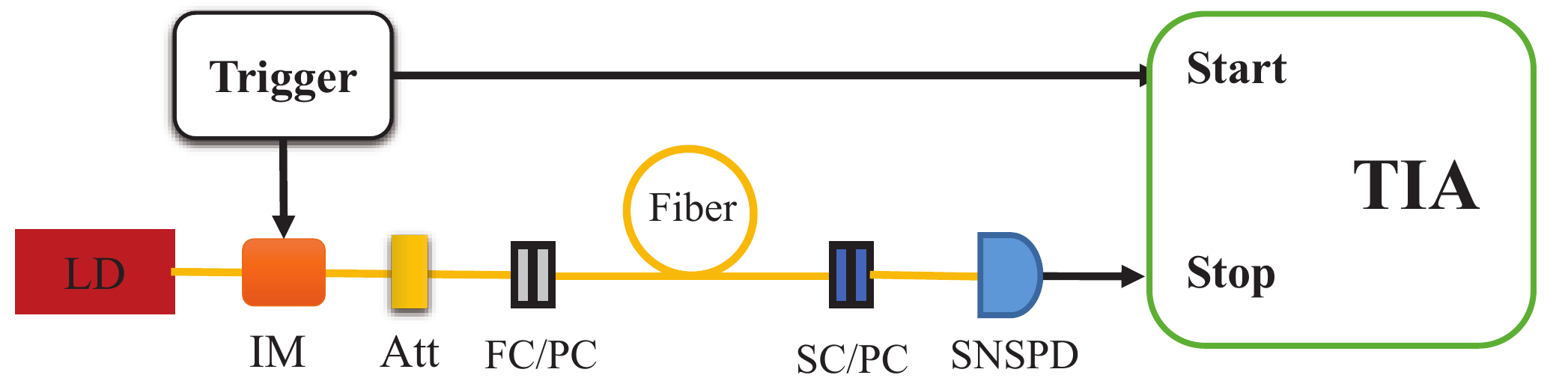}
	\caption{(Color online) setup of the delay measurement. We use a continuous-wave laser diode (LD) at 1560 nm, an intensity modulator (IM), and an attenuator (Att) to generate pulse signals with a temporal width of $\sim100$ ps. Then we acquire the two arrival-time distributions of the photons after travelling through the fiber segments of interest, using an SNSPD and a time interval analysis (TIA, Picoquant HydraHarp400). The delay of the fiber spool is calculated as the difference of the mean values of the two distributions. Since the IM and the start of the TIA are triggered by the same electronics, the same detector is used for both distributions, and we use the time-tag unit, the difference is only affected by the resolution of the TIA, reaching 1 ps.
	}\label{measuredelay}
\end{figure}

The time delay of commercial off-the-shelf fiber spools is measured using a conventional optical time domain reflectometer (OTDR). By itself, it has a low accuracy and does not meet the sensitivity demands of our experiment. For example, the distance uncertainty of a conventional state-of-the-art OTDR (Exfo FTB7600E) is $  \pm (0.75 + 0.001\times L)$ meter, corresponding to a time-delay uncertainty of  $\pm (3.75+0.001\times L/v)$ ns, where  $L$ is the length of the fiber spool and $v$ is group velocity in fiber. In our experiment, we home made all the fiber segments with a time-delay uncertainty of $\sim 2.5$ ps using the following method:
\begin{enumerate}
	\item To make one of the segments whose time delay is $2\times(2^{k-1}+1)$ ns, we cut off a $2\times(2^{k-1}+1)+5$ ns time-delay fiber with an FC/PC connector at one end and an SC bare fiber terminator (Thorlabs) at the other end, using a commercial OTDR to approximate the length. For the segments whose time delay is in the event dead zone of the OTDR, we directly make it by using a ruler with a resolution of 0.5 mm (= 2.5 ps).
	\item We measure the time delay of the segment using our home-built photon-counting OTDR variant shown in Fig.\ \ref{measuredelay}, which achieves a resolution of  1 ps. Then, we cleave the redundant fiber whose length is measured by a ruler with a resolution of 0.5 mm (= 2.5 ps).
	\item We pull the fiber out the bare fiber terminator. Then the fiber is connectorised and is slightly polished with an SC/FC connector achieving a low insertion loss.
\end{enumerate}

The resultant length of each of the fiber segments for Alice and Bob is shown in Table \ref{tabledelay}. For different combinations of fiber segments, the largest deviation of the delay, compared to the target value, is 101 ps for Alice and 46 ps for Bob---both of which are far below 900 ps (the maximum deviation of the delay if we want to apply the correct phase to the photon pulse in the desired time bin).
\begin{table*}[ht!]
	\scriptsize
	\centering
	\caption{ (a) Length and loss of each of Alice's fiber segments. (b) Length and loss of each of Bob's fiber segments. For the $k$-th segment, the length of the long option and that of the short option are $t\times (2^{k-1}+1)$ and $t$, respectively, where $t=2$ ns is the pulse width of phase modulation. Errors shown represent 1 standard deviation. }
	\subtable[]{
		\begin{tabular}{|ccccccc|}
			\hline \hline
			\multicolumn{1}{|c|}{\diagbox{Option}{$k$}}&1&2&3&4&5&6\\
			\hline
			\multirow{1}{*}{Long length~(ns)}&$4.006\pm0.001$ &$ 6.007\pm0.001 $&$ 10.004\pm0.000 $&$18.012\pm0.001 $&$33.999\pm0.002 $ & $66.002
			\pm0.001 $ \\
			
			Loss~(dB)&$0.101\pm0.001$ &$ 0.090\pm0.002 $&$ 0.139\pm0.002 $&$0.115\pm0.002 $&$0.125\pm0.003 $ & $0.130
			\pm0.001 $ \\
			\hline
			\multirow{1}{*}{Short length~(ns)}&$ 2.008 \pm0.001 $ &$ 2.010 \pm0.001 $ & $ 2.008 \pm0.001 $ &$ 2.008 \pm0.001 $ & $ 2.010
			\pm0.001 $ &$ 2.003 \pm0.001 $\\
			Loss~(dB)&$0.104\pm0.002$ &$ 0.078\pm0.001 $&$ 0.068\pm0.003 $&$0.062\pm0.002 $&$0.046\pm0.003 $ & $0.031
			\pm0.003 $ \\
			\hline \hline
			\multicolumn{1}{|c|}{\diagbox{Option}{$k$}}&7&8&9&10&11&12\\
			\hline
			\multirow{1}{*}{Long length~(ns)}&$ 130.008 \pm0.001 $&$257.948
			\pm0.001 $&$513.995\pm0.001$ &$1025.985\pm0.001 $&$ 2049.936\pm0.002 $&$4098.000\pm0.005 $\\
			Loss~(dB)&$0.089\pm0.002$ &$ 0.110\pm0.002 $&$ 0.154\pm0.001 $&$0.145\pm0.001 $&$0.198\pm0.003 $ & $0.198
			\pm0.001 $ \\
			\hline
			\multirow{1}{*}{Short length~(ns)}&$ 2.003 \pm0.001 $ &$2.009 \pm0.001 $ & $ 2.006 \pm0.001 $ &$ 2.009 \pm0.001 $&$ 2.009 \pm0.001 $&$ 2.008 \pm0.001 $\\
			Loss~(dB)&$0.053\pm0.002$ &$ 0.072\pm0.001 $&$ 0.094\pm0.002 $&$0.110\pm0.001 $&$0.082\pm0.003 $ & $0.076
			\pm0.002 $ \\
			\hline \hline
			
		\end{tabular}\label{Tabalice}
	}

	\subtable[]{
		\begin{tabular}{|ccccccc|}
			\hline \hline
			\multicolumn{1}{|c|}{\diagbox{Option}{$k$}}&1&2&3&4&5&6\\
			\hline
			\multirow{1}{*}{Long length~(ns)}&$4.013\pm0.001$ &$6.010\pm0.001 $&$ 10.003\pm0.001$&$18.004\pm0.001$ &$34.009 \pm0.000 $ & $66.056 \pm0.001 $ \\
			Loss~(dB)&$0.0149\pm0.001$ &$0.0176\pm0.001 $&$ 0.250\pm0.001$&$0.098\pm0.002$ &$0.218 \pm0.001 $ & $0.170 \pm0.001 $ \\
			\hline
			\multirow{1}{*}{Short length~(ns)}&$ 2.008 \pm0.001 $ & $ 1.999 \pm0.001 $ & $ 2.008 \pm0.001 $ &$ 2.009 \pm0.001 $ & $ 2.007 \pm0.001 $ &$2.010 \pm0.001 $\\
			Loss~(dB)&$0.073\pm0.002$ &$0.106\pm0.002 $&$ 0.096\pm0.001$&$0.076\pm0.001$ &$0.081 \pm0.001 $ & $0.098 \pm0.001 $ \\
			\hline \hline
			\multicolumn{1}{|c|}{\diagbox{Option}{$k$}}&7&8&9&10&11&12\\
			\hline
			\multirow{1}{*}{Long length~(ns)}&$ 130.002 \pm0.001 $&$257.992 \pm0.001 $&$514.004 \pm0.001$ &$ 1025.999\pm0.002 $&$ 2049.961\pm0.004 $& $ 4097.993
			\pm0.007$\\
			Loss~(dB)&$0.100\pm0.002$ &$0.070\pm0.003 $&$ 0.110\pm0.002$&$0.040\pm0.003$ &$0.400 \pm0.001 $ & $0.280 \pm0.001 $ \\
			\hline
			\multirow{1}{*}{Short length~(ns)}&$ 2.005\pm0.001 $ & $ 2.006 \pm0.001 $ & $ 2.005 \pm0.001 $ &$ 2.004\pm0.001 $ & $ 2.007 \pm0.001 $ &$ 2.002\pm0.001 $\\
			length~(ns)&$0.089\pm0.001$ &$0.043\pm0.002 $&$ 0.073\pm0.002$&$0.033\pm0.003$ &$0.128 \pm0.001 $ & $0.113 \pm0.001 $ \\
			\hline\hline
			
		\end{tabular} \label{Tabbob}
	}
	\label{tabledelay}
\end{table*}

\section{Detailed Experimental results}\label{detailed experimental result}
\begin{figure}[!htbp]
	\subfigure[]{\label{error11bits}
		\centering
		\includegraphics[width=0.63\linewidth]{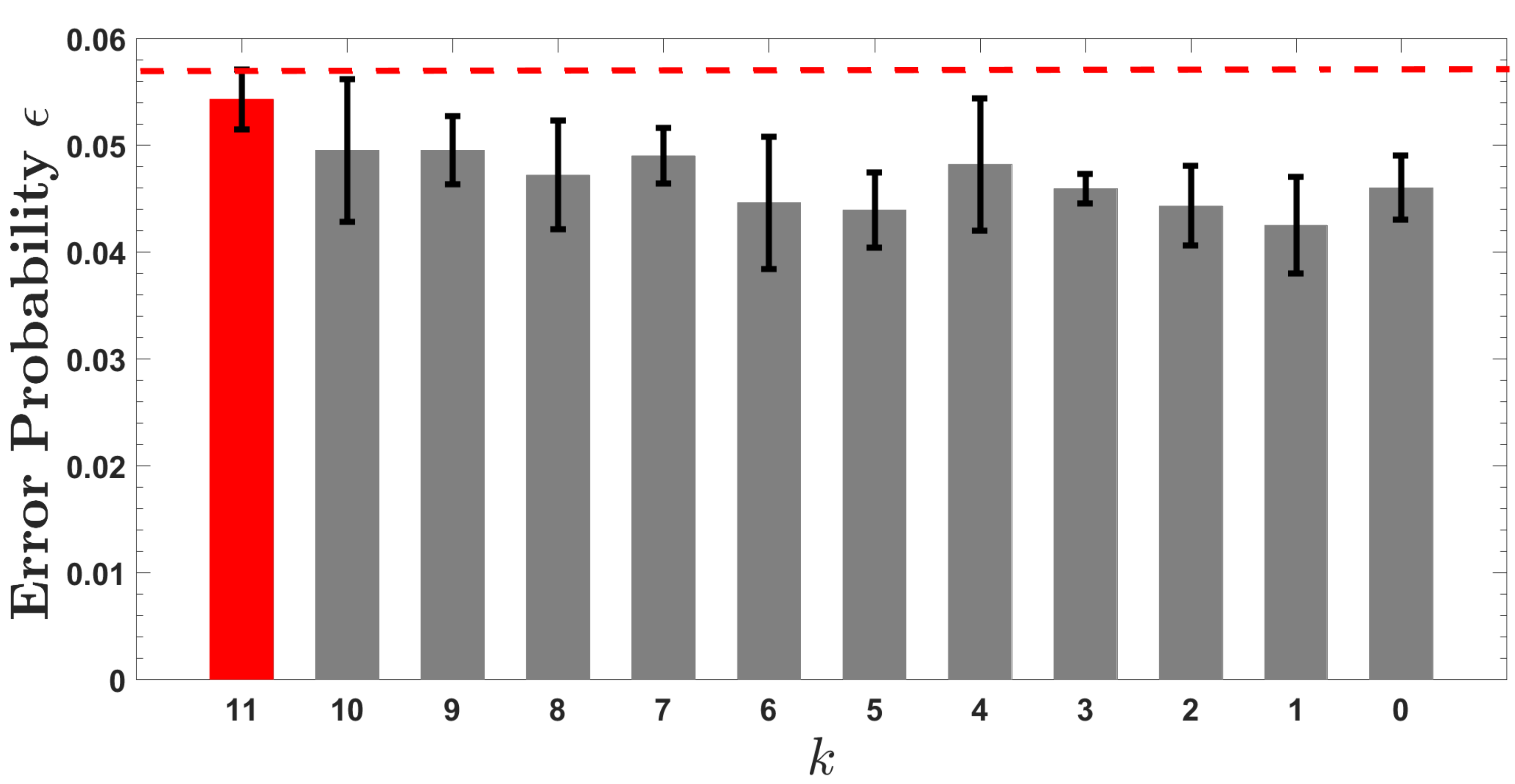}
		
	}
	\subfigure[]{\label{error10bits}
		\centering
		\includegraphics[width=0.63\linewidth]{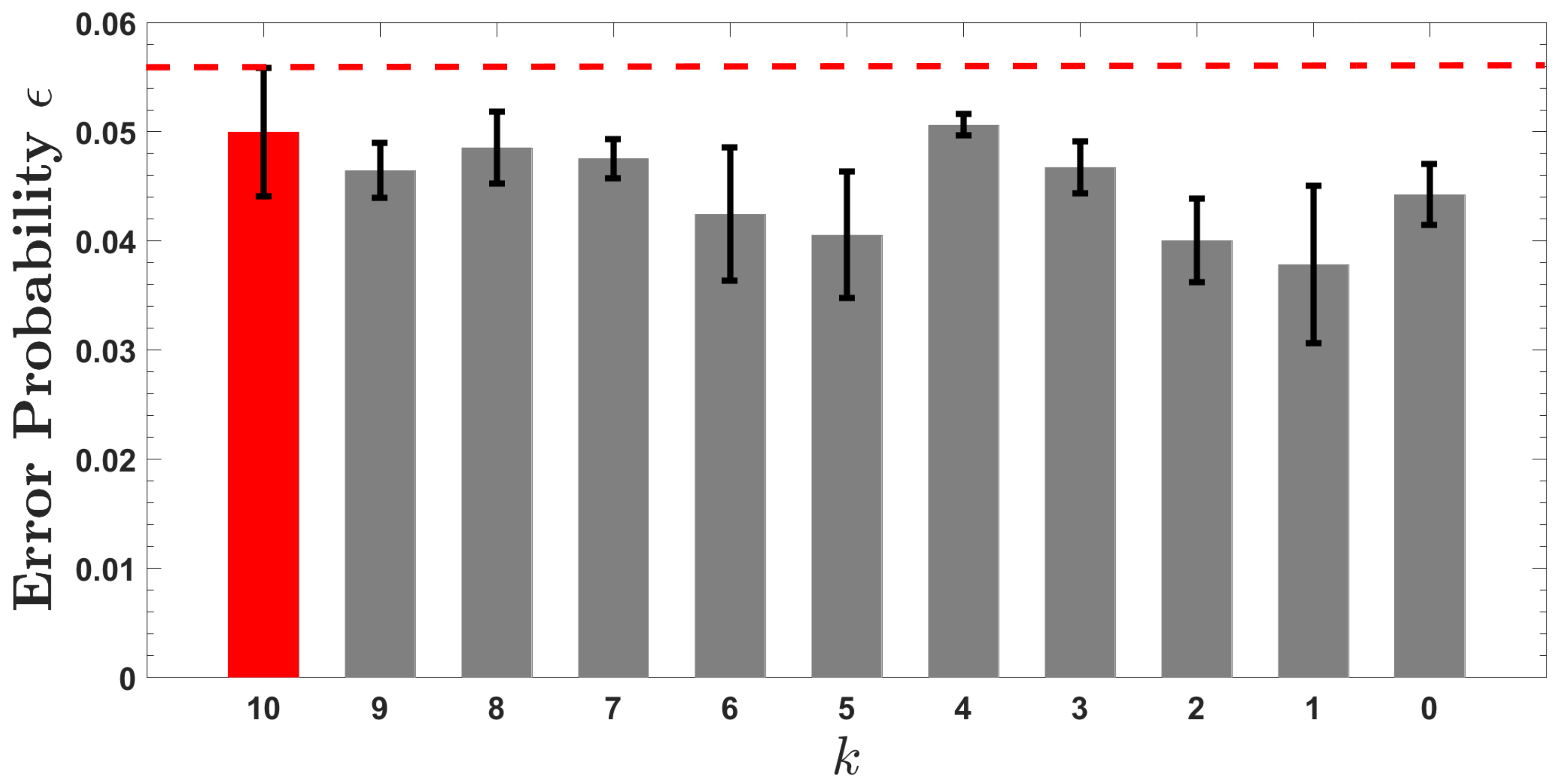}
		
	}
	\subfigure[]{\label{error9bits}
		\centering
		\includegraphics[width=0.63\linewidth]{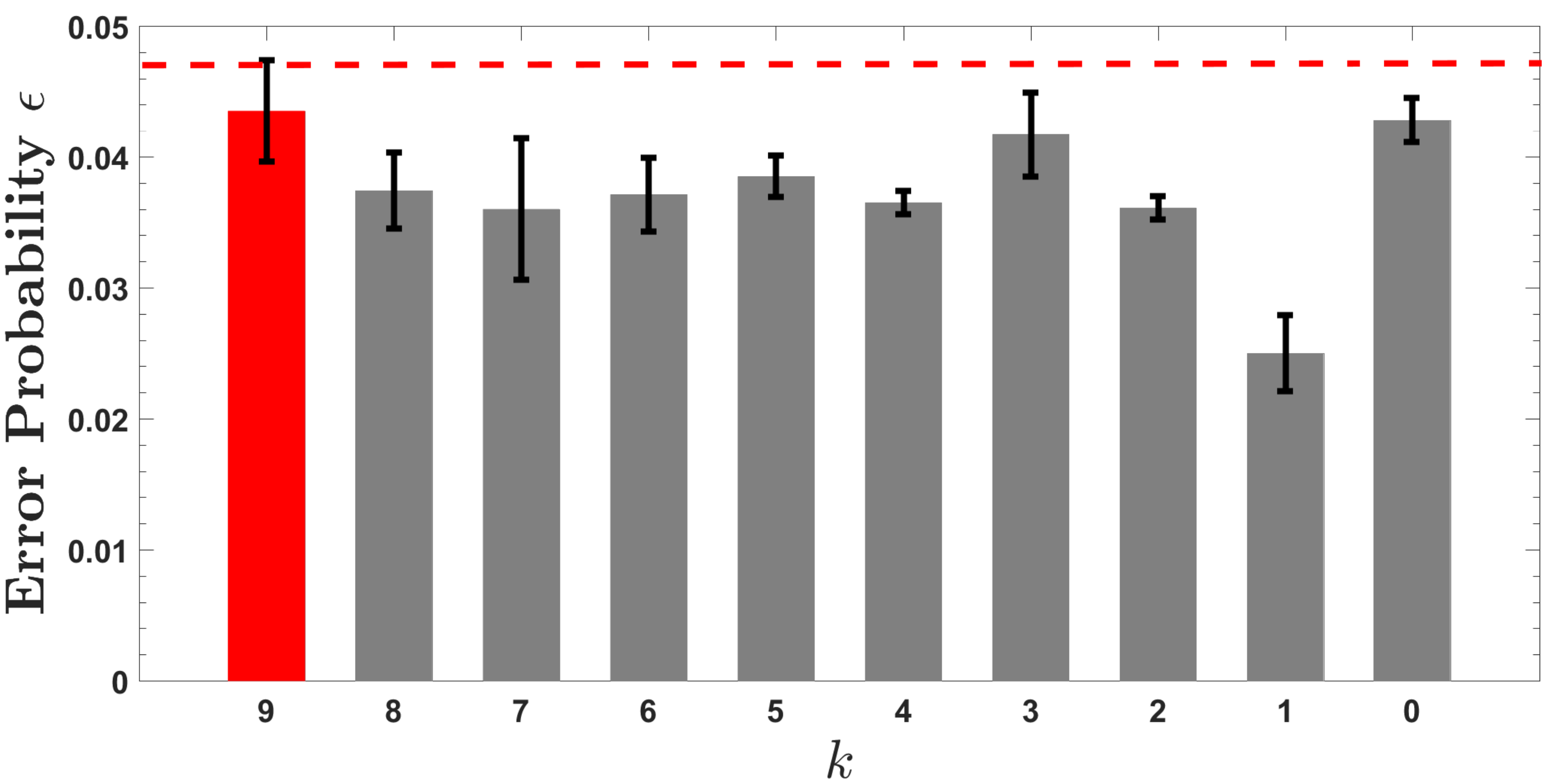}
		
	}
	\caption{ (a) Error probability for the system size $n=11$. (b) Error probability for the system size $n=10$. (c) Error probability for the system size $n=9$. For each system size $n$, the red bar denotes the worst case of inputs $x=y=2^{n}-1$. The other values of $k\in{\{0,1,\ldots,11\}}$ correspond to Alice's $(k+1)$-th bit being flipped from 1 to 0. The red dashed line illustrates the worst error probability accounting for the uncertainty. The error bars indicate 1 standard deviation.}\label{Figerror}
\end{figure}

Table \ref{tableresults} shows the complete experimental results.
Figure\ \ref{Figerror} shows the experimental error probabilities of different system sizes ranging from $n=11$ to 9. For each system size, the largest error probability is obtained in the worst case (red bar). The error bars represent 1 standard deviation.  A possible reason why the error rises with growing system size is because the thermal expansion introduced by setting the different length of each section cannot be compensated, and results in a length mismatch as $n$ grows. \blk
\begin{table*}[ht!]
	\caption{ Detailed experimental results. The system loss and the error probability are estimated for the worst-case inputs for Alice and Bob (consisting of $x=y=2^n-1$ and $f(z)=g(z)=0$ for $z$ even and $f(z)=g(z)=1$ for $z$ odd). The error bars indicate 1 standard deviation.}
	%\resizebox{\linewidth}
	\begin{tabular}{ccccc}
		\hline \hline
		System size ($n$)&9&10&11&12\\
		Loss of Alice's section~(dB)&1.34$\pm$0.04&1.39$\pm$0.04&1.78$\pm$0.04&2.06$\pm$0.04\\
		Loss of Bob's section~(dB)&0.9$\pm$1.06&1.26$\pm$0.04&1.26$\pm$0.04&1.46$\pm$0.04\\
		System loss (dB)&$13.86\pm0.04$&$14.06\pm0.04$&$14.66\pm0.04$&$15.14\pm0.04$\\
		
		Error probability ($\epsilon$)&$0.0405\pm0.0023$&$0.0498\pm0.0059$&$0.0543\pm0.0028$&$0.0638\pm0.0025$\\
		
		Q&$271.26\pm2.50$&$308.45\pm2.84$&$383.66\pm3.53$&$461.45\pm4.25$\\
		
		$\gamma$ (classical)&$1.636\pm0.015$&$0.930\pm0.008$&$0.578\pm0.005$&$0.348\pm0.003$\\
		
		$\gamma$ (quantum definite)&$3.272\pm0.030$&$1.860\pm0.017$&$1.157\pm0.010$&$0.696\pm0.006$\\
		
		\hline  \hline
	\end{tabular}
	\label{tableresults}
\end{table*}


%\newpage
% \clearpage
%\onecolumngrid

\end{document}